\documentclass{amsart}
\usepackage{graphicx}
\usepackage{amscd}

\newtheorem{theorem}{Theorem}
\theoremstyle{plain}

\newtheorem{corollary}{Corollary}

\newtheorem{lemma}{Lemma}

\newtheorem{remark}{Remark}

\numberwithin{equation}{section}

\begin{document}
\title[Behavior of De Haan-Renick estimator]{THE WEAK LIMITING BEHAVIOR OF THE DE HAAN RESNICK ESTIMATOR
OF\ THE EXPONENT OF A STABLE DISTRIBUTION}

\author{Gane Samb LO}

\address{Lo Gane Samb, Universit\'{e} Paris 6. L.S.T.A. Tour 45-46, E.3., 4, Place Jussieu. 75230, Paris Cedex 05.}

\begin{abstract}
\large
The problem of estimating the exponent of a stable law received a
considerable attention in the recent literature. Here, we deal with an
estimate of such a exponent introduced by De Haan and Resnick when the
corresponding distribution function belongs to the Gumbel's domain of
attraction. This study permits to construct new statistical tests. Examples and simulations are given. The limiting law
are shown to be the Gumbel's law and particular cases are given with norming
constants expressed with iterated logarithms and exponentials.
\end{abstract}

\maketitle

\Large

\noindent \textbf{Nota Bena}. This paper is an unpublished of the author, part of his PhD thesis, UPMC, Paris VI, 1986.\\

\section{INTRODUCTION AND RESULTS} \label{sec1}

\noindent Many biological phenomena seem to fit the Zipf's form :

\begin{equation}
1-G\left( x\right) =C\text{ }x^{-1/c},\text{ }c>0\text{ and }C>0.
\label{1.1}
\end{equation}

\noindent For instance, we can cite the plot against r of the population of the r-th
largest city (see e.g. Hill \cite{hill75}. This motivated considerable works on
the problem of estimating c. More generally, if $X_{1},X_{2},...,X_{n}$ are
independent and identical copies of a random variable (rv) X such that $%
F\left( x\right) =\mathbb{P}\left( X\leq x\right) $ satisfies 

\begin{equation}
\forall \text{ }t>0,\text{ \ \ \ }\lim_{x\uparrow +\infty }\frac{1-F\left(
\log \left( tx\right) \right) }{1-F\left( \log \left( x\right) \right) }%
=t^{-1/c}.  \label{Ac}
\end{equation}

\noindent Several estimates (Hill \cite{hill75}, S. Csorgo-Deheuvels-Mason \cite{cdm} have been
proposed. De Haan and Resnick \cite{haan-resnick} introduced 

\begin{equation*}
T_{n}=\left( X_{n,n}-X_{n-k,n}\right) /\log k,
\end{equation*}

\noindent where $X_{1,n},X_{2,n},...,X_{n,n}$ are the order statistics of $%
X_{1},X_{2},...,X_{n}$ and k is a sequence of integers satisfying

\begin{equation*}
0<k<n,\text{ \ \ \ }k=k\left( n\right) \rightarrow +\infty ,\text{ \ \ }%
k/n\rightarrow 0\text{ as }n\rightarrow +\infty \tag{K}
\end{equation*}

\noindent De Haan and Resnick \cite{haan-resnick} have proved that (\ref{Ac}) implies under some
conditions that :

\bigskip 
\begin{equation}
T_{n}\rightarrow ^{P}c,\text{ in probabiblity \ }\left( \rightarrow
^{P}\right)  \label{i}
\end{equation}

\noindent and 
\begin{equation}
\frac{\log k}{c}\left( T_{n}-c\right) \rightarrow \Lambda \text{ , in
distribution }\left( \rightarrow ^{d}\right)  \label{ii}
\end{equation}

\noindent where 

\begin{equation*}
\Lambda \left( x\right) =\exp \left( e^{-x}\right) \text{ }
\end{equation*}

\noindent is the Gumbel law.\\

\noindent In order to contribute to a complete asymptotic theory for the inference
about the upper tail of a distribution (as specified in LO \cite{loa}, Section
3), we deal with the asymptotic behavior of $T_{n}$, here, in the case where (%
\ref{Ac}) fails. Notice that (\ref{Ac}) mean that $F\left( \log \left(
.\right) \right) $ belongs to the Frechet's domain of attraction. Here, we
restrict ourselves to the case where $F\left( \log \left( .\right) \right) $
belongs to the Domain of attraction of the Gumbel law. $D\left( \Lambda
\right) $. These results are stated in this section \ref{sec1}, proved in section \ref{sec2}
and illustrated in Section \ref{sec3}.\\

\noindent Before the statements of the results, we need some further notation. Define 

\begin{equation*}
A=\inf \left\{ x,\text{ }F\left( x\right) =1\right\} \text{ },\text{ }B=\sup
\left\{ x,\text{ }F\left( x\right) =0\right\},
\end{equation*}

\begin{equation*}
R\left( t\right) =\left( 1-F\left( t\right) \right) ^{-1}\int_{t}^{A}\left(
1-F\left( v\right) \right) \text{ }dv,\text{ }B\leq t<A
\end{equation*}

\noindent and

\begin{equation*}
Q\left( u\right) =F^{-1}\left( u\right) =\inf \left\{ x,\text{ }F\left(
x\right) \geq u\right\}
\end{equation*}

\noindent is the quantile function of X. We shall assume, when appropriate, that 
\begin{equation}
F\left( \log \left( x\right) \right) \in D\left( \Lambda \right).  \tag{H1}
\end{equation}

\begin{equation}
F\left( x\right) \text{ is ultimately strictly increasing and continuous}.
\tag{H2}
\end{equation}

\noindent We will prove that of $(H1)$ and $(H2)$ are satisfied, then $F\in
D\left( \Lambda \right) $ (see Lemma \ref{l2}). So, \ we can use the De Haan (see \cite{haan})
representation for the quantile function associated with a distribution
function F such that $F\in D\left( \Lambda \right) $ and $F\left( x\right) $
is ultimately strictly increasing as $x\uparrow A$ (see De Haan \cite{haan},
Theorems 1.4.1 and 2.4.2) : 

\begin{equation}
Q\left( 1-u\right) =c_{o}+r\left( u\right) +\int_{u}^{1}\left( r\left(
t\right) /t\right) \text{ }dt\text{, as }u\downarrow 0.  \label{1.2}
\end{equation}

\noindent where $c_{o}$ is some constant anf $r\left( u\right) $ is a positive
function slowly varying at zero (S.V.Z).\\

\noindent Finally, we define this assumption of $k$. We say that $k$ satisfies $(Kr\left(\lambda \right) )$ and its
satisties $(K)$ and this extra-condition :
\begin{equation*}
\ r\left( 1/n\right) /r\left( k/n\right) \rightarrow \lambda ,0\leq \lambda
<+\infty
\end{equation*}

\noindent Our man results are

\begin{theorem} Let $(H1)$ and $(H2)$ be satisfied, then

\noindent (i) for any sequence $k$ satisfying $(K)$, we have 

\begin{equation}
\left( X_{n,n}-X_{n-k,n}\right) /\log k\text{ \ }\rightarrow ^{p}0 \label{1.3}
\end{equation}

\noindent (ii) for any sequence k satisfying $(Kr\left( \lambda \right) )$, we have 

\begin{equation}
\frac{\log k}{r\left( k/n\right) }\left\{ \frac{X_{n,n}-X_{n-k,n}}{\log k}%
\right\} -\frac{Q\left( 1-1/n\right) -Q\left( 1-k/n\right) }{r\left(
k/n\right) } \rightarrow ^{d} \lambda \times \Lambda  \label{1.4}
\end{equation}

\noindent with $a_{N} =r\left( k/n\right)$ and $b_{n}=\left\{ Q\left( 1-1/n\right)
-Q\left( 1-k/n\right) \right\} \text{ }/\text{ }r\left( k/n\right)$.
\end{theorem}

\bigskip

\begin{corollary} \label{c1}
Let $A>0$. Then, if $(H1)$ and $(H2)$ hold, (\ref{1.3}) and (\ref
{1.4}) remain true if we replace 
\begin{equation*}
X_{n,n},\text{ }X_{n-k,n}\text{ by }\log X_{n,n},\text{ }\log X_{n-k,n}
\end{equation*}

\begin{equation*}
Q\left( 1-s\right) \text{ by }\log Q\left( 1-as\right) ,\text{ }a=P\left(
X>0\right)
\end{equation*}

\begin{equation*}
r\left( s\right) \text{ by }s\left( u\right) =R\left( Q\left( 1-u\right)
\right) /Q\left( 1-u\right)
\end{equation*}

\noindent and if k satisfies $Ks\left( \lambda \right) $ at the place of $Kr\left(
\lambda \right) $. Moreover, if $\log $ $A>0$, we may repeat the operation. Conversely
\end{corollary}

\begin{corollary} \label{c2}
Let $r\left( u\right) \rightarrow 0$ as $u\downarrow 0$. Then, if $(H1)$
and $(H2)$ are satisfied, (\ref{1.3}) and (\ref{1.4}) hold if we replace

\begin{equation*}
X_{n,n},\text{ }X_{n-k,n}\text{ by }\exp \left( X_{n,n}\right) ,\text{ }\exp
\left( X_{n-k,n}\right)
\end{equation*}

\begin{equation*}
Q\left( 1-u\right) \text{ by }\exp \left( Q\left( 1-u\right) \right)
\end{equation*}

\begin{equation*}
r\left( u\right) \text{ by }t\left( u\right) =\exp \left( Q\left( 1-u\right)
\right) \text{ }R\left( Q\left( 1-u\right) \right)
\end{equation*}

\noindent and if k satisfies $Kt\left( \lambda \right) $ at the place of $Kr\left(
\lambda \right) $.
\end{corollary}

\begin{corollary} \label{c3} (Particular cases). Here, we restrict ourselves to the case where $k=[ \left( \log n\right)
^{\ell }]$, where $[\cdot]$ denotes the integer part, and $\ell $ is any positive
number.\\

\noindent 1) \textbf{Normal case} : $X\sim \mathcal{N}\left( 0,1\right) $\\

\noindent (i) 
\begin{equation*}
\ell \left( 2\log n\right) ^{\frac{1}{2}}\left( \log \log n\right) \text{. }%
T_{n}.\text{ }\left( 1+o\left( 1\right) \right) -\ell \left( \log \log
n\right) \left( 1+o\left( 1\right) \right) \rightarrow ^{d}\Lambda
\end{equation*}

\noindent (ii) 

\begin{equation*}
\left( 2\log n\right) ^{\frac{1}{2}}T_{n}\rightarrow ^{P}1
\end{equation*}

\noindent 2) \textbf{Exponential case} : $\exp \left( X\right) \sim \mathcal{E}\left( 1\right) .$ (or
general gamma case, see the proof).\\

\noindent (i) 
\begin{equation*}
\left( 1+o\left( 1\right) \right) .\text{ }\ell \left( \log n\right) \left(
\log \log n\right) \text{ }T_{n}-\ell \left( \log \log n\right) \left(
1+o\left( 1\right) \right) \rightarrow ^{d}\Lambda
\end{equation*}

\noindent (ii) 

\begin{equation*}
\left( \log n\right) \text{ }T_{n}\rightarrow ^{P}1
\end{equation*}

\noindent  3) Suppose that $X=\log _{p}\sup \left( e_{p-1}\left( 1\right),Z\right)$, $a.s.$, where $Z\sim \mathcal{N}\left( 0,1\right)$,  $p\geq 1$, $\log _{p}$ (resp. $e_{p}$) denotes the p-th logarithm (resp. exponential), with by convention $\log _{0}x=1$,for all $x>0$. Let 

\begin{equation*}
C_{n}\sim \left( 2\log n\right) \prod_{h=1}^{h=p-1}\log _{h}\left( 2\log
n\right) ^{\frac{1}{2}}.
\end{equation*}

\noindent Then we have :\\

\noindent (i) First,

\begin{equation*}
\ell \left( \log \log n\right) \text{ }C_{n}\text{ }T_{n}-\ell \left( \log
\log n\right) \left( 1+o\left( 1\right) \right) \rightarrow ^{d}\Lambda;
\end{equation*}

\noindent (ii) Nnxt,

\begin{equation*}
C_{n}\text{ }T_{n}\rightarrow ^{P}1.
\end{equation*}
\end{corollary}

\begin{remark} \label{r1}
\bigskip  Mason \cite{mason} has proved that the Hill (\cite{hill75})estimate 

\begin{equation*}
H_{n}=k^{-1}\sum_{i=1}^{i=k}\left( X_{n-i+1,n}-X_{n-k,n}\right)
\end{equation*}

\noindent is characteristic of a distribution satisfying (\ref{Ac}) in the following
sense : suppose that $A=+\infty $, then for any real number c, $0<c<+\infty $%
, one has 

\begin{equation*}
T_{n}\rightarrow ^{P}c,\text{ for any sequence k satisfying (K)}
\end{equation*}

\noindent if and only if F satisfies (\ref{Ac}). In order to compare $H_{n}$ and $T_{n}$, we remark that this property is not obtained for $T_{n}$. Indeed, the Mason distribution difined by 

\begin{equation*}
F^{-1}\left( 1-2^{-m}\right) =m,\text{ }m=0,1,2,... \\
F^{-1}\left( 1-u\right)
\end{equation*}

\begin{equation*}
 =m+\left( 2^{-m}-u\right) 2^{m+1},\text{ if }%
2^{-m-1}<u<2^{-m}
\end{equation*}

\noindent satisfies the following property 

\begin{equation*}
\frac{F^{-1}\left( 1-s\right) -F^{-1}\left( 1-bs\right) }{\log b}+\left(
\log 2\right) ^{-1}\text{ as }s\rightarrow 0,\text{ }b\rightarrow +\infty 
\text{ and }bs\rightarrow 0.
\end{equation*}

\noindent Thus, by letting $b=U_{k,n}.U_{1,n}^{-1}$ $,$ $s=U_{1,n},$ we get $%
T_{n}\rightarrow ^{P}\left( \log 2\right) ^{-1}$, where we have used the
well known representation $X_{n-i+1,n}$ $=^{d}$ $F^{-1}\left(
U_{n-i+1,n}\right) $ $d_{=}$ $F^{-1}\left( 1-U_{i,n}\right) $ and $%
U_{1,n}\leq U_{2,n}\leq ...\leq U_{n,n}$ are the order statistics of a
sequence of independent rv's uniformly distributed on $\left( 0,1\right) $.
However, $1-F\left( \log \left( x\right) \right) $ does not vary regulary at
infinity, in other words, does not satisfy (\ref{Ac}), $c=\left( \log
2\right) ^{-1}$. (see e.g. Mason (1982), Appendix).
\end{remark}

\bigskip

\section{Proofs of the Results} \label{sec2} 
\noindent First, we show how to derive corollaries \ref{c1} and \ref{c2} from the Theorems. To begin
with, we need four lemmas and we define $\mathcal{L}$ as the set of distribution functions
F satisfying $(H1)$ and $(H2)$

\bigskip

\begin{lemma} \label{l1} If $F\in \mathcal{L}$, then (\ref{1.2}) holds and $\lim_{u\downarrow 0}\frac{r\left(
u\right) }{R\left( Q\left( 1-u\right) \right) }=1$. It follows that $R\left(
Q\left( 1-u\right) \right) $ is S.V.Z.
\end{lemma}

\begin{lemma} \label{l2} Let $A>0$. Then, if X has a distribution function $F\in L$, $%
\log \sup \left( 0,X\right) $ is defined a.s., and has a distribution
function $G\in L$ and $S\left( G^{-1}\left( 1-u\right) \right) \sim \frac{%
R\left( Q\left( 1-u\right) \right) }{Q\left( 1-u\right) },$ as $u\downarrow
0 $, where $S\left( t\right) =\int_{t}^{\log A}\frac{1-G\left( v\right) }{%
1-G\left( t\right) }dv,$ $-\infty <t<\log A$
\end{lemma}

\begin{lemma} \label{l3} Let $R\left( t\right) \rightarrow 0$ as $t\uparrow 0$. Then, if X has a
distribution function $F\in L$, $\exp \left( X\right) $ has a distribution
function H such that $H\in L$ and $R\left( Q\left( 1-u\right) \right) \sim 
\frac{T\left( H^{-1}\left( 1-u\right) \right) }{H^{-1}\left( 1-u\right) }$
as $u\downarrow 0$, where $T\left( t\right) =\left( 1-H\left( t\right)
\right) ^{-1}\int_{t}^{e^{A}}\left( 1-H\left( v\right) \right) $ $dv$, $%
e^{B}<t<e^{A}$
\end{lemma}

\begin{lemma} \label{l4} If $E\in L$, then $Q\left( 1-u\right) $ is S.V.Z.
\end{lemma}

\bigskip

\begin{proof} Proofs of Lemmas \ref{l1}, \ref{l2}, \ref{l3} and \ref{l4}.\\

\noindent Lemmas \ref{l1}, \ref{l2} and \ref{l3} are proved in Lo \cite{lob}  via lemma 3.2, 3.3 and 3.4. Lemma \ref{l4} is proved in Lo \cite{loa} via its lemma 2.\\

\noindent \textbf{Proof of corallaries \ref{c1} and \ref{c2}}.\\

\noindent Let $G$ be the distribution function of $\log \sup \left( 0,X\right) $. Lemma
\ref{l2} says that $(H1)$ and $(H2)$ imply that $F\in D\left( \Lambda
\right) $. But $g\left( \log x\right) =F_{1}\left( x\right) $, where $F_{1}$
is the distribution function if $\sup \left( 0,X\right) $. And it is obvious
that $F_{1}\in D\left( \Lambda \right) $ if $A>0$. If follows that $(H1)$ and $(H2)$ are true for $G$. So, we may write (\ref{1.3}) and (\ref{1.4}%
) for G. Furthermore, we have $G^{-1}\left( 1-u\right) =\log Q\left(
1-as\right) ,$ $a=P\left( X>0\right) $ and lemmas 1.2 say that we may
replace $r\left( u\right) \sim R\left( Q\left( 1-u\right) \right) $ by $%
S\left( G^{-1}\left( 1-u\right) \right) \sim \frac{R\left( Q\left(
1-u\right) \right) }{Q\left( 1-u\right) }$. Finally, remark that if $A>0$, $k$
satisfies $(K)$, we have for large values of $n$, $\sup \left(
0,X_{n-k,n}\right) =X_{n-k,n}$ and $\sup \left( 0,X_{n,n}\right) =X_{n,n},$ $%
a.s.$. With the above remarks, we can see that Corolarry 1 is proved.\\

\noindent \textbf{Corollary \ref{c2}} is proved a similar way with Lemma \ref{l3}.\\

\noindent \textbf{Proof of the part (i) of the theorem}. Let $G\left( x\right) =F\left( \log x\right) $. Since $G\in L$, lemma \ref{l4} implies that $G^{-1}\left( 1-u\right) $ is S.V.Z. At this step, we need the
Karamata's representation for functions S.V.Z. 

\begin{equation}
G^{-1}\left( 1-u\right) =c\left( u\right) \exp \left( \int_{u}^{1}\left(
\varepsilon \left( s\right) /s\right) \text{ }ds\right),  \label{2.1}
\end{equation}

\noindent where $ ,\text{ }c\left(s\right) \rightarrow c,\text{ }0<c<+\infty ,\text{ }\varepsilon \left(
s\right) \rightarrow 0\text{ as }s\downarrow 0$. So, 

\begin{equation}
Q\left( 1-u\right) =\log G^{-1}\left( 1-u\right) =\log c\left( u\right)
+\int_{u}^{1}\left( \varepsilon \left( s\right) /s\right) \text{ }ds,
\label{2.2}
\end{equation}

\noindent We recall that 

\begin{equation}
\left\{ X_{i,n}\text{ },\text{ }1\leq i\leq n\right\} \text{ }d_{=}\text{ }%
\left\{ Q\left( U_{i,n}\right) ,\text{ }1\leq i\leq 1\right\} ,U_{i,n}\text{ 
}d_{=}\text{ }1-U_{n-i+1,n},\label{2.3}
\end{equation}

\noindent where $0=U_{0,n}\leq U_{1,n}\leq U_{2,n}\leq ...\leq U_{n,n}\leq U_{n+1,n}=1$ are
the order statistics of a sequence of independent rv's uniformly distributed
on (0,1). Therefore, (\ref{2.3}) implies 

\begin{equation}
\left( \log k\right) \text{ }T_{n}=X_{n,n}-X_{n-k,n}\text{ }d_{=}\text{ }%
Q\left( 1-U_{1,n}\right) -Q\left( 1-U_{k-1,n}\right) =\left( \log k\right) 
\text{ }T_{n}^{\ast }.  \label{2.4}
\end{equation}

\noindent Let us apply (\ref{2.3}). We obtain that 

\begin{equation*}
0 \leq T_{n}^{\ast }\leq \frac{\left| \log c\left( U_{1,n}\right) /c\left(
U_{k-1,n}\right) \right| }{\log k}+\frac{\left| \log \left(
nU_{k+1,n}\right) \right| }{\log k}\sup_{0\leq s\leq U_{k+1,n}}\left|
\varepsilon \left( s\right) \right|   
\end{equation*}

\begin{equation}
\leq :A_{n1}+A_{n2}. \label{2.5}
\end{equation}

\noindent Obviously, we have 

\begin{equation}
A_{n1}\rightarrow ^{P},\text{ since }U_{1,n}\rightarrow ^{P}0\text{ and }%
U_{k+1,n}\rightarrow ^{P}0\text{ if k satisfies $(K)$.}  \label{2.6}
\end{equation}

\noindent By (\ref{2.1}), we also have that 

\begin{equation}
\sup_{0\leq s\leq U_{k+1,n}}\left| \varepsilon \left( s\right) \right|
=o_{p}\left( 1\right)  \label{2.7}
\end{equation}

\noindent Therefore, we can see that (\ref{2.5}), (\ref{2.6}) and (\ref{2.7}) will
imply the part (i) of theorem 2 if we prove that $\left( \log
nU_{k+1,n}\right) /\log k=0_{p}\left( 1\right) $. But (see e.g. De Haan and
Balkema \cite{balkema}),

\begin{equation}
nk^{-\frac{1}{2}}\left( U_{k,n}-\frac{k}{n}\right) \rightarrow ^{d}N\left(
0,1\right) ,\text{ which implies}  \label{2.8}
\end{equation}

\begin{equation}
n\left( k+1\right) ^{-1}U_{k+1,n}\rightarrow ^{P}1.  \label{2.9}
\end{equation}

\noindent We deduce from (\ref{2.9}) that 

\begin{equation}
\left( nk^{-1}U_{k+1,n}\right) \rightarrow ^{P}1. \label{2.10}
\end{equation}

\noindent So,

\begin{equation}
\left( \log nU_{k+1,n}\right) /\log k=0_{p}\left( 1\right),  \label{2.10b}
\end{equation}

\noindent which completes the proof of the part (i) of the theorem.\\

\noindent \textbf{Proof of the part (ii) of the theorem}. Let suppose that $(H1)$ and $(H2)$ hold. Then, (\ref{1.2}) holds. 

\begin{equation}
Q\left( 1-u\right) =c_{o}+r\left( u\right) +\int_{u}^{1}\left( r\left(
s\right) /s\right) \text{ }ds\text{, as }u\downarrow 0.
\end{equation}

\noindent We have 

\begin{equation*}
T_{n}^{\ast } =\frac{\left\{ Q\left( 1-U_{1,n}\right) -Q\left(
1-1/n\right) \right\} }{\log k}+\frac{\left\{ Q\left( 1-k/n\right) -Q\left(
1-U_{k+1,n}\right) \right\} }{\log k} \\
\end{equation*}

\begin{equation*}
+\frac{a_{n}.b_{n}}{\log k}.
\end{equation*}

\begin{equation}
=:A_{n3}/\log k+A_{n4}/\log k+a_{n}.b_{n}/\log k,  \label{2.11}
\end{equation}

\noindent where $a_{n}$ and $b_{n}$ are defined in the statement of the theorem. First, we prove that 

\begin{equation}
A_{n4}/a_{n}\rightarrow ^{P}0  \label{2.12}
\end{equation}

\noindent By (\ref{1.2}), we have $A_{n4}=r\left( k/n\right) -r\left( U_{k+1,n}\right)
+\int_{k/n}^{U_{k+1,n}}\left( r\left( s\right) /s\right) $ $ds$.\\

\noindent Remark that since $r\left( u\right) $ is slowly varying at 0, we have on
account of (\ref{2.10}) that 

\begin{equation*}
\frac{r\left( k/n\right) }{r\left( U_{k+1,n}\right) }\rightarrow 1\text{, in
probability}.
\end{equation*}

\noindent Furthermore, 

\begin{equation}
a_{n}^{-1}\left| \int_{k/n}^{U_{k+1,n}}\left( r\left( s\right) /s\right) 
\text{ }ds\right| \leq \left| \log \left( nk^{-1}U_{k+1,n}\right) \right|
\sup_{s\in I_{n}}\left| \frac{r\left( s\right) }{r\left( k/n\right) }\right|,
\label{2.14}
\end{equation}

\noindent where 
\begin{equation*}
I_{n}=\left\{ \min \left( \frac{k}{n},\text{ }U_{k+1,n}\right) ,\text{ }\max
\left( \frac{k}{n},\text{ }U_{k+1,n}\right) \right\}
\end{equation*}

\noindent is a random interval. At this step, we need this

\begin{lemma} \label{l5}
Let $r\left( u\right) $ be a positive function S.V.Z. Let $\left(
u_{n}\right) $ be the sequence of rv's and $\left( d_{n}\right) $ be a
sequences of real numbers such that 

\begin{equation*}
d_{n}\rightarrow +\infty ,\text{ }u_{n}\rightarrow ^{P}0,\text{ }\left(
d_{n}.u_{n}\right) =0_{p}\left( 1\right) ,\text{ and }\left(
d_{n}.u_{n}\right) ^{-1}=0_{p}\left( 1\right) \text{ as }n\rightarrow +\infty
\end{equation*}

\noindent then 
\begin{equation*}
1+o_{p}\left( 1\right) =inf_{s\in J_{n}}\frac{r\left( s\right) }{r\left(
1/d_{n}\right) }\leq \sup_{s\in J_{n}}\frac{r\left( s\right) }{r\left(
1/d_{n}\right) }=1+o_{p}\left( 1\right) ,\text{ as }n\rightarrow +\infty
\end{equation*}

\noindent where 
\begin{equation*}
J_{n}=\left( \min \left( \frac{1}{d_{n}},u_{n}\right) ,\text{ }\max \left( 
\frac{1}{d_{n}},u_{n}\right) \right).
\end{equation*}
\end{lemma}

\noindent \textbf{Proof of Lemma \ref{l5}} The proof is the same as that of Lemma 3.5 in Lo (1985b).\\

\noindent \textbf{Proof of the part (ii) of the theorem (continued)}.\\
\\
\noindent By (\ref{2.10b}), $\frac{k}{n}U_{k+1,n}\rightarrow ^{P}1.$ So, we may apply
lemma \ref{l5} to (\ref{2.14}) and get 

\begin{equation}
a_{n}^{-1}\left| \int_{k/n}^{U_{k+1,n}}\left( r\left( s\right) /s\right) 
\text{ }ds\right| \leq o_{p}\left( 1\right) .\left( 1+o_{p}\left( 1\right)
\right)  \label{2.15}
\end{equation}

\noindent Combining (\ref{2.12}) and (\ref{2.15}), we get (\ref{2.12}). We now concentrate on $A_{n3}$ and show that 

\begin{equation}
A_{n3}/a_{n}\rightarrow ^{d}\lambda \times \Lambda,  \label{2.16}
\end{equation}

\noindent whenever if k satisfies $\left(Kr\left(\lambda \right)\right)$. We have 

\begin{equation*}
A_{n3}=Q\left( 1-U_{1,n}\right) -Q\left( 1-1/n\right)
\end{equation*}

\begin{equation}
=r\left( U_{1,n}\right) -r\left( 1/n\right) -\int_{U_{1,n}}^{1/n}\left(
r\left( s\right) /s\right) \text{ }ds.  \label{2.17}
\end{equation}

\noindent Recall that 

\begin{equation}
P\left( n\text{ }U_{1,n}\geq x\right) \rightarrow e^{-x},\text{ as }%
n\rightarrow +\infty.  \label{2.18}
\end{equation}

\noindent This means that $nU_{1,n}=0_{p}\left( 1\right) $ and $\left( nU_{1,n}\right)
^{-1}=0_{p}\left( 1\right) $. Thus, we may apply Lemma \ref{l5} and get 

\begin{equation}
r\left( 1/n\right) /r\left( U_{1,n}\right) \rightarrow ^{P}1\text{ as }%
n\rightarrow +\infty.  \label{2.19}
\end{equation}

\noindent Then, if $k$ satisfies $\left( Kr\left( \lambda \right) \right) $, we have 
\begin{equation}
\left( r\left( 1/n\right) -r\left( U_{1,n}\right) \right) /a_{n}\rightarrow
^{P}0\text{, as }n\uparrow +\infty.  \label{2.20}
\end{equation}

\noindent Now, let 
\begin{equation*}
B_{n}=\int_{U_{1,n}}^{1/n}\left( r\left( s\right) /s\right) \text{ }ds.
\end{equation*}

\noindent We have 
\begin{equation*}
B_{n}/a_{n}=\left\{ r\left( 1/n\right) /r\left( k/n\right) \right\} .\text{ }%
\int_{U_{1,n}}^{1/n}\left\{ \frac{r\left( s\right) }{r\left( 1/n\right) }%
\right\} \frac{ds}{s}.
\end{equation*}

\noindent Then, it follows, from Lemma \ref{l5} and the fact that $r\left( u\right) $ is
positive, that if $k$ satisfies $\left( Kr\left( \lambda \right) \right) $, we
have 

\begin{equation}
\left\{ B_{n}/a_{n}\right\} +\lambda \log nU_{1,n}=\log nU_{1,n}.\text{ }%
o_{p}\left( 1\right).  \label{2.21}
\end{equation}

\noindent By (\ref{2.18}), we see that 
\begin{equation*}
\lim_{n\uparrow +\infty }P\left( -\log nU_{1,n}\leq x\right)
=e^{-e^{-x}}=\Lambda \left( x\right).
\end{equation*}

\noindent We get finally that if k satisfies $\left( Kr\left( \lambda \right) \right) $%
, one has 

\begin{equation}
B_{n}/a_{n}\rightarrow ^{d}\lambda .\Lambda  \label{2.30}
\end{equation}

\noindent (\ref{2.16}) and (\ref{2.12}) together imply the theorem.\\

\noindent \textbf{Proof of the corollary \ref{c3}}.\\

\noindent Previously in the occasion of our study of the same particular cases for the
Hill's estimator (see Lo \cite{loa}), Lemma 5 and Corollary 5) we have proved
that (\ref{3.1}) 
\begin{equation}
\lim_{u\downarrow 0}R\left( Q\left( 1-u\right) \right)
/\rho \left( u\right) \rightarrow 1, \label{3.1}
\end{equation}

\noindent where $\rho \left( u\right) =u$,  $Q^{\prime }\left( 1-u\right)$, $Q^{\prime }\left( u\right) =\frac{dQ\left(
u\right) }{du}$ for values of u near 1.\\

\noindent With (\ref{3.1}), we may handle the different points of Corollary \ref{c3}. Here,
we concentrate on the case where $k=\left( \left( \log n\right) ^{\ell
}\right) ,$ $\ell >0$.\\

\noindent 1) - \textbf{Normal case} : $F\left( x\right) =\int_{-\infty }^{x}\left( 2\pi \right)
^{-\frac{1}{2}}e^{\frac{-t^{2}}{2}}dt$.\\

\noindent Remark that $F\left( \log x\right) $ is the distribution function of the
log-normal law. It follows that $F\left(\log\left( .\right) \right) \in
D\left( \Lambda \right) $. $(H2)$ is obviously true. On the other hand,
it is well known that

\begin{equation}
Q\left( 1-s\right) =\left( 2\log \left( 1/s\right) \right) ^{\frac{1}{2}}+%
\frac{\log \log \left( 1/s\right) +4\pi +o\left( 1\right) }{2\left( 2\log
\left( 1/s\right) \right) },\text{ as }s\downarrow 0  \label{3.2}
\end{equation}

\begin{equation}
\rho \left( s\right) =s\text{ }Q^{\prime }\left( 1-s\right) =\left( 2\log
\left( 1/s\right) \right) ^{-\frac{1}{2}}\left( 1+o\left( 1\right) \right) ,%
\text{ as }s\downarrow 0.  \label{3.3}
\end{equation}

\noindent Notice that we might have used (see Galambos \cite{galambos}, p. 66)

\begin{equation*}
R\left( t\right) =t^{-1}\left( 1+o\left( t^{-3}\right) \right) \text{ and }%
\rho \left( s\right) \sim R\left( Q\left( 1-s\right) \right). 
\end{equation*}

\noindent Then

\begin{equation*}
a_{n}^{-1}=\rho ^{-1}\left( k/n\right) \left( 1+o\left( 1\right) \right)
=\left( 2\log n\right) ^{\frac{1}{2}}\left( 1+o\left( 1\right) \right) 
\end{equation*}

\noindent and

\begin{equation*}
b_{n}=Q\left( 1-1/n\right) -Q\left( 1-k/n\right) =\frac{\ell \log \log
n}{\left( 2\log n\right) ^{\frac{1}{2}}}\left( 1+0\left( \frac{\log \log n}{%
\log n}\right) \right) 
\end{equation*}

\noindent Therefore,
\begin{equation*}
b_{n}=\left\{ \ell \log \log n\right\} \left( 1+o\left( 1\right) \right) 
\end{equation*}

\noindent 2) \textbf{Exponential case} : $F\left( \log x\right) =1-e^{-\alpha x},$ $\alpha >0$.

\noindent More generally, since the tail of the quantile function associated with a
general gamma law $\gamma \left( r,\alpha \right) ,$ $r>0$, $\alpha >0$,
admits the expansion

\begin{equation}
H^{-1}\left( 1-u\right) =\left( \log \frac{1}{u}\right) \left( 1+o\left(
1\right) \right)   \label{3.4}
\end{equation}

\noindent the behavior of $T_{n}$ is same for all Gamma laws because (\ref{3.4})
doesn't depend neither on $r$,  nor on $\alpha $. That is why we only consider

\begin{equation*}
F\left( \log x\right) =1-e^{-x}.
\end{equation*}

\noindent Therefore

\begin{equation*}
Q\left( 1-u\right) =\log \log \left( 1/s\right) \text{, }\rho \left(
u\right) =\left( \log \left( 1/s\right) \right) ^{-1}.
\end{equation*}

\noindent Then
\begin{equation*}
a_{n}=\left( \log n\right) \left( 1+o\left( 1\right) \right) ,\text{ }%
b_{n}=\left( \ell \log \log n\right) \left( 1+o\left( 1\right) \right) 
\end{equation*}

\noindent At this step, we apply the Theorem to conlude.\\

\noindent 3) In this case, we have

\begin{equation*}
T_{n}=\frac{\left\{ \log _{p}Z_{n,n}-\log _{p}Z_{n-k,n}\right\} }{\log k},
\end{equation*}

\noindent for large values of n, where $Z_{1,n},Z_{2,n},...,Z_{n,n}$ are the order
statistics of a sequence of independent and standard Gaussian rv's.\\

\noindent We also have

\begin{equation}
1-G\left( x\right) =m^{-1}\left( \int_{x}^{+\infty }\left( 2\pi \right) ^{-%
\frac{1}{2}}e^{-\frac{t^{2}}{d}}dt\right) ,\text{ }  \label{3.6}
\end{equation}

\noindent where G is the distribution function associated to $\sup \left(
e_{p-1}\left( 1\right) ,\text{ }Z\right) ,$ and $m=p\left( Z>e_{p-1}\left(
1\right) \right)$.\\

\noindent Since $X=\log \sup \left( e_{p-1}\left( 1\right) ,\text{ }Z\right) ,$ one has

\begin{equation*}
Q\left( 1-u\right) =\log _{p}G^{-1}\left( 1-mu\right) =\log _{p}\left\{
\left( 2\log \left( 1/ms\right) \right) ^{\frac{1}{2}}+\frac{\log \log
\left( 1/ms\right) +0\left( 1\right) }{2\left( 2\log \left( 1/ms\right)
\right) ^{\frac{1}{2}}}\right\}. 
\end{equation*}

\noindent It follows from (\ref{3.3}) and (\ref{3.6}) that

\begin{equation*}
\rho ^{-1}\left( s\right) =\left( 2\log \left( 1/s\right) \right)
\prod_{j=1}^{j=p-1}\log _{j}\left( 2\log \left( 1/s\right) \right) ^{\frac{1%
}{2}}\left( 1+o\left( 1\right) \right) 
\end{equation*}

\noindent Then

\begin{equation*}
a_{n}\sim 2\log n\prod_{j=p}^{j=p-1}\log _{j}n
\end{equation*}

\noindent and from (\ref{3.6}), we deduce after some calculations that

\begin{equation*}
b_{n}=\ell \log \log n\left( 1+o\left( 1\right) \right). 
\end{equation*}

\noindent Remark that the part 3 of the corollary might have been derived from the
part 1 of the same corollary after p applications of Corollary \ref{c1}. We have
given the normal case as example but such an operation is possible whenever $%
Z_{i}$ has a distribution function $F$ such that $F\left( \log \left( .\right)
\right) \in D\left( \Lambda \right) $ and $\log _{p-1}A>0$. Even when $F\left(
\log \left( .\right) \right) \in D\left( \Psi \right) $, where $\Psi \left(
x\right) =e^{-1/x}$ is the Frechet law, we can have the part 3 since $%
F\left( \log \left( .\right) \right) \in D\left( \Psi \right) $ implies that 
$F\left( .\right) \in D\left( \Lambda \right) .$
\end{proof}

\newpage

\section{Simulations} \label{sec3}

\noindent Here, we will illustrate the behavior of $T_{n}$ in the three cases.\\

\noindent (i) $\exp \left( X\right) \sim E\left( 1\right) $, (ii) $\exp \left(
X\right) =\sup \left( 0,Z\right) ,$ $Z\sim N\left( 0,1\right) ,$ (iii) $%
F\left( \log x\right) =1-1/x$.\\

\noindent For making our simulations, we have generated an ordered sample $u_{i},$ $%
1\leq i\leq 4000$ from a uniform rv. Therefore, we have constructed :\\

\noindent (i) an order sample of the standard exponential law

\begin{equation*}
y_{i}=-\log \left( 1-u_{i}\right) 
\end{equation*}

\noindent and defined $T_{n1}=\left( \log y_{n}-\log y_{n-k}\right) /\log k$,  $k=[n^{\frac{1}{2}}]$, $T_{n1}=C_{n}\left( y_{i}\right)$;\\

\noindent (ii) an ordered sample of the standard Normal law for $u_{i}\uparrow 1$%

\begin{equation*}
x_{i}=\left( -2\log \left( 1-u_{i}\right) \right) ^{\frac{1}{2}}\text{ (see
e.g. 3.2)}
\end{equation*}

\noindent and defined

\begin{equation*}
T_{n2}=C_{n}\left( x_{i}\right); 
\end{equation*}

\noindent (iii) an ordered sample of the Pareto law

\begin{equation*}
z_{i}=\left( 1-u_{i}\right) ^{-1}\text{ and difined }T_{n3}=C_{n}\left(
z_{i}\right) 
\end{equation*}

\noindent Before we proceed any further, we remark that with $x_{i}=\left( -2\log
\left( 1-u_{i}\right) \right) $, we get:

$T_{n2}=\frac{1}{2}T_{n1}$. In fact, we have $T_{n2}^{\ast }=\frac{1}{2}%
T_{n1}$%

\begin{equation*}
+0\left( \frac{\log \log \left( 1/u_{n-k,n}\right) }{4\log \text{ }k\text{ }%
\log \left( 1/1-u_{n-k,n}\right) }\right) 
\end{equation*}

\noindent where $T_{n2}^{\ast }$ is the exact value of $T_{n2}$ if we use the true
quantile function. With our data, we get $T_{n2}^{\ast }=\frac{1}{2}%
T_{n1}^{-1}\pm 0.0179$ and $\left( \log n\right) T_{n2}^{\ast }=\left( \frac{%
1}{2}\log n\right) T_{n1}\pm 0.14$.\\

\noindent The simulations are given as follows :

\subsection{The data} \label{subsec31}

\noindent Here are the simulation outcomes.

\begin{center}
\begin{tabular}{|c|c|c|c|c|}
\hline
   $n$ & $(\frac{1}{2}\log n)T_{n1}$  & $(\log n)T_{n2}$  & $T_{n3}$ & $u_{4000-i+1}$  \\
\hline
3991 & 0.3294 & 0.3294 & 0.3302  & 0.002435  \\
\hline
3992 & 0.3391 & 0.3391  & 0.4042 & 0.001631\\
\hline              
3993 & 0.3625 & 0.3625  & 0.4334  & 0.002620 \\
\hline
3994 & 0.3550 & 0.3550 & 0.4324   & 0.001337 \\              
\hline
3995 & 0.4598 & 0.4598 & 0.5954   & 0.000988\\
\hline
3996 & 0.4693 & 0.4593 & 0.6124 & 0.000437 \\
\hline              
3997 & 0.4689 & 0.4689 & 0.6130 & 0.000418 \\
\hline
3998 & 0.4977 & 0.4977 & 0.6625 & 0.000308\\
\hline
3999 &  0.5116& 0.5116 & 0.6963 & 0.000297\\
\hline
4000 & 0.6942 & 0.9942 & 1.00332 & 0.000095 \\
\hline
1          &    2      &   3     &   4    &  5      \\
\hline
\end{tabular}
\end{center}

\subsection{Comments} \label{subsec32} We make few comments about the simulations.\\

\noindent 1) The right column gives values of $\left( u_{i}\right)$, first. With the
symmetry of the uniform maw, we have 

$$
\left\{ 1-u_{i},1\leq i\leq 4000\right\} d_{=}\left\{ u_{4000-i+1},1\leq i\leq 4000\right\}
$$,

\noindent 2) If $k=[n^{\frac{1}{2}}]$, similar calculations as in the
proof of the part 2 of corollary \ref{c3} show that $a_{n}\sim \log n$ and $%
a_{n}.b_{n}\sim \log 2.$ Therefore, if $\bar{T}_{n1}$ denotes the De
Haan/Resnick estimate for $\exp \left( X\right) \sim E\left( 1\right) ,$ we
get

\begin{equation}
\left( \frac{1}{2}\log n\right) \bar{T}_{n1}\rightarrow ^{P}\log 2.
\label{4.1}
\end{equation}

\noindent The same considerations from part 3 of corollary \ref{c3} $\left( p=1\right) $
yield 

\begin{equation}
\left( \log n\right) \bar{T}_{n2}\rightarrow ^{P}\log 2,  \label{4.2}
\end{equation}

\noindent where $\bar{T}_{n2}$ denotes the De Haan/Resnick estimate for $X=\log \sup
\left( 0,Z\right) ,$ $Z\sim \left( 0,1\right) $. Notice that (\ref{4.1}) and
(\ref{4.2}) are well illustrated by our simulations since $\log 2\sim 0.69314
$.\\

\noindent 3) The column 4 illustrates the result of De Haan/Resnick (1980)

\begin{equation*}
T_{n3}\rightarrow^{P}1.
\end{equation*}

\section{Conclusions} \label{sec4}

\noindent We have proved that a suitable choice of $k$ (for instance $k\sim \left\{ \log
n\right\} ^{\ell }$), we can find the norming constants $d_{n}$ such that

\begin{equation*}
d_{n}.T_{n}\rightarrow ^{P}1
\end{equation*}

\noindent In addition, we have given the limit law as the Gumbel distribution. The
same work has been already done by De Haan and Resnick (1980) under the
hypothesis (\ref{Ac}). So, for a wide range of distributions belonging in $D\left(
\Lambda \right) $, we can provide statistical tests. For instance,
we may obtain tests for a Normal model against an Exponential one based on (\ref{4.1}) and (\ref{4.2}). Similar tests are specified in LO (1985a)

\end{document}